\documentstyle[11pt,epsf]{article}

\newcommand{\bea}{\begin{eqnarray}}
\newcommand{\eea}{\end{eqnarray}}

\def\void{}
\def\labelmark{}
\newenvironment{formula}[1]{\def\labelname{#1}
\ifx\void\labelname\def\junk{\begin{displaymath}}
\else\def\junk{\begin{equation}\label{\labelname}}\fi\junk}%
{\ifx\void\labelname\def\junk{\end{displaymath}}
\else\def\junk{\end{equation}}\fi\junk\labelmark\def\labelname{}}
{\ifx\void\labelname\def\junk{\end{array}\end{displaymath}}
\else\def\junk{\end{array}\right.\end{equation}}
\fi\junk\labelmark\def\labelname{}\def\junk{}
}
\newcommand{\beq}{\begin{formula}}
\newcommand{\eeq}{\end{formula}}
\newcommand{\beqv}{\begin{formula}{}}
\begin{document}

\begin{titlepage}

\null
\begin{flushright}
SU-4240-616\\
\end{flushright}
\vspace{10mm}

\begin{center}
\bf\Large Baby Universes in 4{\it d}\\
          Dynamical Triangulation 
\end{center}

\vspace{5mm}

\begin{center}
{\bf S. Catterall}\\
Physics Department, Syracuse University,\\
Syracuse, NY 13244.\\
{\bf J. Kogut}\\
Loomis Laboratory, University of Illinois at Urbana,\\
1110 W. Green St, Urbana, IL 61801.\\
{\bf R. Renken}\\
Department of Physics, University of Central Florida,\\
Orlando, FL 32816. \\
{\bf G. Thorleifsson}\\
Physics Department, Syracuse University,\\
Syracuse, NY 13244.
\end{center}

\begin{center}
\today
\end{center}

\vspace{5mm}
      
\begin{abstract}
We measure numerically the distribution of baby universes
in the crumpled phase of the dynamical triangulation model
of 4{\it d} quantum gravity.
The relevance of the results to the issue of an exponential
bound is discussed. The data are consistent with the
existence of such a bound.
\end{abstract}

\vfill

\end{titlepage}

One of the more promising approaches to understanding the nature
of four-dimensional quantum gravity has arisen through models
based on summing classes of simplicial manifolds -
the dynamical triangulation (DT) models \cite{blanket,ambrev}.
The manifold is approximated by a set of equilateral simplices
whose edge lengths are taken to constitute an invariant cutoff.
Quantum fluctuations of the geometry are incorporated by
constructing a partition function which sums over all 
possible ways of assembling these simplices into a piecewise
linear manifold
\begin{equation}
Z\left(\kappa_0,\kappa_4\right)=\sum_{T\left(S^4\right)}e^{\kappa_0 N_0
-\kappa_4 N_4} \;.
\end{equation}
Here the class of triangulations has been restricted to that
of spherical topology. 
The coupling $\kappa_4$ constitutes a bare cosmological constant
conjugate to the total number of four-simplices (volume) $N_4$.
Similarly,
$\kappa_0$ plays the role of an inverse Newton constant coupled
to the total number of zero-simplices (nodes) $N_0$ in the
triangulation.

Assuming that we wish to remove the edge length cutoff it is necessary
to find points in the parameter space of the model at which the
mean volume $\left<N_4\right>$ diverges. To see how this may
happen consider expanding the {\it grand canonical} partition
function Eq.\ (1) as a power series in $e^{-\kappa_4}$,
\begin{equation}
Z\left(\kappa_0,\kappa_4\right)=\sum_{N_4}\Omega\left(\kappa_0,
N_4\right)e^{-
\kappa_4 N_4}\;.
\end{equation}
The coefficients in this expansion are the microcanonical
partition functions for the system 
at fixed volume $N_4$. It is these quantities
which are estimated in Monte Carlo simulations. In two dimensions
it is known rigorously that the analogous coefficients behave
as $\Omega(N_2)\sim e^{\kappa_2^c N_2}$ - that is there
is an exponential bound on the number of triangulations composed
of $N_2$ triangles provided we restrict 
the global topology sufficiently\footnote{In two 
dimensions this restriction amounts to fixing the genus of the
surface.}.
The existence of this bound ensures that the 
expansion has a finite radius of convergence determined by the
critical coupling~$\kappa_2^c$. The mean volume can then
be shown to diverge in power-like fashion 
as this critical coupling is
approached. This is the basis for taking the continuum limit.

In dimensions greater than two the volume
dependence of $\Omega(\kappa_0,
N_4)$
even when restricted to the four sphere is, in principle, unknown.
In a previous paper we pointed out that the behaviour for
small volume is consistent with a super-exponential growth
$\Omega\left(N_4\right)\sim e^{\beta N_4\log{N_4}}$ \cite{usbound}.
This would, at least naively, render a continuum limit impossible.
Since then two other groups have examined the issue on
larger lattices and claim strong evidence for an exponential
bound \cite{ambbound,enzo}. In light of this we have both 
extended our 
calculations to larger volumes and in addition looked at 
alternative quantities such as the distribution of baby
universes.  The latter is very sensitive to the volume 
dependence of $\Omega(\kappa_0,N_4)$ and might thus be
useful in resolving this issue. 

\begin{figure}
\begin{center}
\leavevmode
\epsfxsize=400pt
\epsfbox{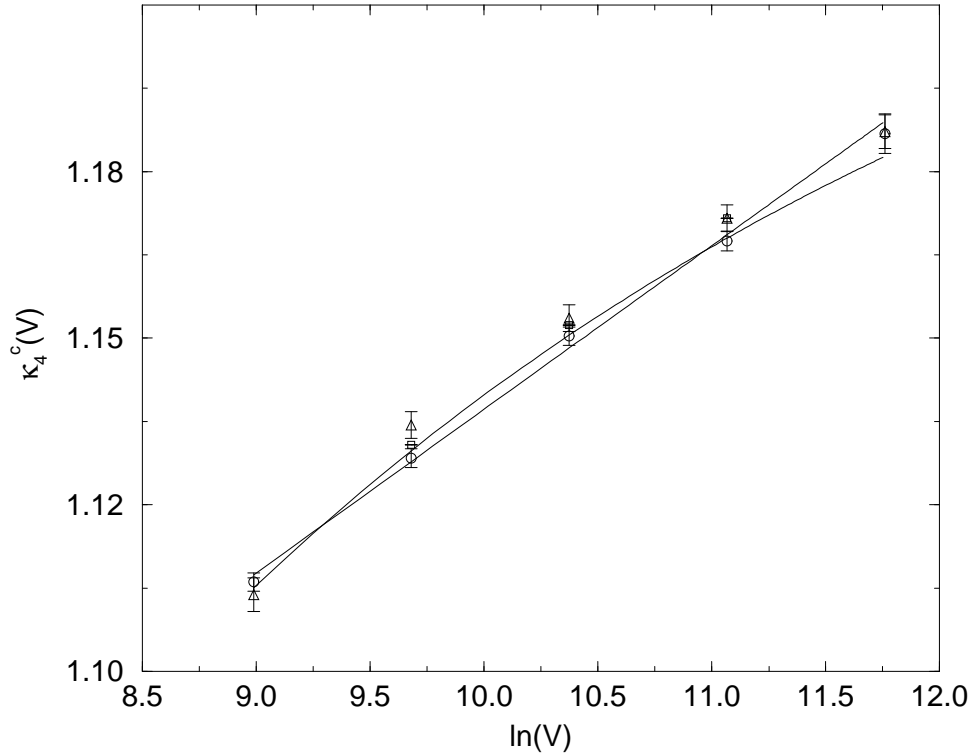}
\caption{The critical coupling $\kappa_4^c(V)$ together with
 fits assuming power-law convergence (curve) and
 super-exponential growth (straight line).}
\end{center}
\end{figure}

The usual way in which an exponential bound is observed
is by looking at the volume dependence of the quantity
$\kappa_4^c(\kappa_0,V)={1\over V}
\log{\Omega(V,\kappa_0)}$
which is a numerical estimate for
the effective critical cosmological coupling $\kappa_4^c$
at volume $V$. 
In Fig.\ 1 
we plot it as a function of the logarithm 
of the volume to expose any logarithmically divergent component to
the critical coupling. 
We show both our data
(circles) together with the data published in \cite{ambbound,enzo}.
Clearly, these different simulations are in agreement within
statistical errors. We then fit 
our data from volumes $V=8000-128000$ using two different 
functional forms.
The straight line represents a least square fit
to a super-exponential form

\begin{equation}
\kappa_4^c(V)=\alpha+\beta\log{V} \;,
\end{equation}
while the curve corresponds to a weak power-law convergence to
an exponential bound of the form

\begin{equation}
\kappa_4^c(V)=\alpha^\prime+{\beta^\prime\over 
V^\gamma}\;.
\end{equation}

\begin{table}
\begin{center}
\begin{tabular}{||l|l|l|c||}\hline
Fit&$\alpha\left(\alpha^\prime\right)$&$\beta\left(\beta^\prime\right)$&
$\chi^2/d.o.f$ \\ \hline
Power $\gamma=0.25$ & 1.242(3) & -1.23(4) & 2.4/3 \\ \hline
Power $\gamma=0.1$  & 1.389(8) & -0.68(2) & 0.8/3 \\ \hline
Log                 & 0.894(8) & 0.025(1) & 3.1/3 \\ \hline
\end{tabular}
\end{center}
\caption{The optimal fit parameters as 
 the data for $\kappa_4^c(V)$
 is fitted to either super-exponential behaviour or
 a weak power-law convergence.}
\end{table}

The fit parameters and quality of the fits are shown in Table 1.
Since there is so little data we have chosen to do the power
fit with two different {\it fixed} powers $\gamma=0.25$ and $\gamma=0.1$.
Arguments for the former choice are made in \cite{ambbound} and it
corresponds to the curve plotted in Fig.\ 1.
It is clear that both types of fit can equally well describe the data. 
The quality of the fit with $\gamma=0.1$ appears somewhat superior 
but since the log fit has $\chi^2$ of order one this
should not be taken as significant. In \cite{enzo}
a $\gamma=0.25$ fit over the same volume range was claimed to
be substantially better than the logarithm. Our data do
not seem to support this and we interpret this as
simply pointing to the delicacy of deciding between similar
fits with rather limited data. 
It is quite possible
that many runs at intermediate volumes would be useful to
resolve this issue.

Thus while we see that an exponential bound is certainly
consistent with the existing numerical data at large volumes it is
not {\it strongly} preferred over the logarithmic divergence.
In light of this we have turned to an analysis of other
quantities to try to settle the issue. The distribution of
baby universes is one such observable \cite{sanjay}. 
A baby universe is defined 
as a section of a {\it d}-dimensional triangulation 
connected to the bulk only through a so-called minimal neck which
consists of $d+1$ $(d-1)$-simplices or faces
constituting a boundary of a simplex {\it not}
already present in the triangulation.
In four dimensions this is a set of five tetrahedral faces
which make up the surface of a new simplex and divide
the triangulation into two pieces. The volume of the baby universe is
defined to be the number of simplices in the smaller piece.

The distribution of these baby universes can be computed by
considering the number of ways a volume $V$ triangulation can be
built from a baby of volume $B$ and a mother of size $V-B$ by
attaching the baby to the mother at some point. This 
gluing operation is effected by identifying one simplex in
the baby with another on the mother. Thus the distribution takes
the form

\begin{equation}
P(B)\;=\; {(V-B)\;\Omega(\kappa_0,V-B)\;B\;\Omega(\kappa_0,B)\over
             \Omega(\kappa_0,V)}\;.
\end{equation}
Strictly speaking the factors $\Omega(\kappa_0,B)$ should be
replaced with one point functions but we shall ignore this
unimportant technicality here. The important thing to notice is that
any exponential factor in $\Omega$ cancels out in this formula
and $P(B)$ only depends on {\it sub-leading} corrections - that is
it is maximally sensitive to the finite volume corrections to
coefficients $\Omega(\kappa_0,V)$.

With this in mind we have measured the distribution $P(B)$
numerically in the crumpled phase of the model when $\kappa_0=0$.
The true volume of the triangulation space is most easily
estimated here since all triangulations contribute with
equal weight to the partition sum. Indeed we do not believe it
is safe to try to estimate the behaviour of $\Omega(\kappa_0,V)$ from
simulations at large $\kappa_0$. At such node couplings the
dominant triangulations correspond to branched polymers whose
mean node number varies linearly with volume. Such configurations
are known to possess an exponential bound. The crumpled
configurations which predominate
at small $\kappa_0$ in contrast have mean node numbers scaling as
some fractional power of the volume. At large $\kappa_0$ 
these latter configurations
will receive large (as $V\to\infty$) exponential suppression
relative to the
branched polymers from the node term in the action. 

We have simulated the model at four different volumes; 500, 1000,
4000 and 8000 simplices, using runs of length $10$ million
sweeps\footnote{One sweep corresponds to $V$ attempted
elementary local moves.}. We will see that the measured distribution
falls off exponentially fast which necessitated such high statistics
runs. This precluded the use of significantly larger lattice
volumes in this study.
Using our previous parameterizations of the finite volume
corrections to $\Omega(V)$ we have attempted
to fit the data with functional forms
corresponding to either logarithmic divergence or weak power
law convergence
\begin{eqnarray}
\log{P\left(B\right)}&=&a+\beta\left(\left(B+\delta\right)
\log{\left(B+\delta\right)}+\left(
V-B+\delta\right)\log{\left(V-B+\delta\right)}\right),\\
\log{P\left(B\right)}&=&a^\prime+
\beta^\prime\left(\left(B+\delta\right)^{1-\gamma}+
\left(V-B+\delta\right)^{1-\gamma}\right).
\end{eqnarray}
The constant $\delta$ is inserted as a phenomenological parameter to
reflect sub-leading finite size corrections and $a$ and $a^\prime$
reflect an ambiguity in overall normalization. In practice we
have removed the largest contribution to the latter by
dividing the measured number of baby universes by the volume $V$.

\begin{figure}
\begin{center}
\leavevmode
\epsfxsize=400pt
\epsfbox{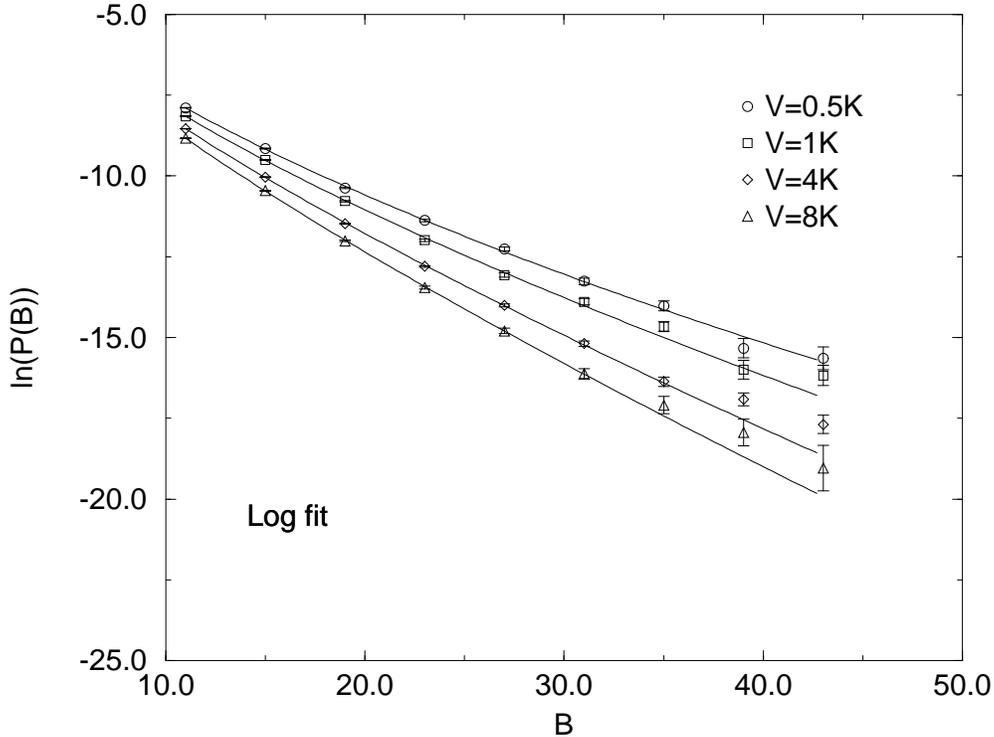}
\caption{$\log{P(B)}$ vs $B$ with a logarithmic fit.}
\end{center}
\end{figure}

Figure 2 shows the distributions 
together with a series of curves resulting
from least-square fits assuming the logarithmic
scenario Eq.\ (6). 
The fit to the largest volume yields $a=-2.92(3)$, $\beta=+0.056(1)$
and $\delta=-7(1)$ with $\chi^2=9.6/6$ (per d.o.f.).
Fits to the other volumes give consistent results.  
Notice that we have fitted baby universes with size $B=4\left(n+1\right)$
only ($n$ integer). Baby universes of size $B=4\left(n-1\right)$
lie on a curve which while yielding consistent fits for $\beta$
is shifted by a constant with respect to the first. This 
effect has been observed before \cite{amb} and is 
presumably the result of finite size effects. We fit only for
$B>10$ and truncate due to poor statistics at $B>50$.

\begin{figure}
\begin{center}
\leavevmode
\epsfxsize=400pt
\epsfbox{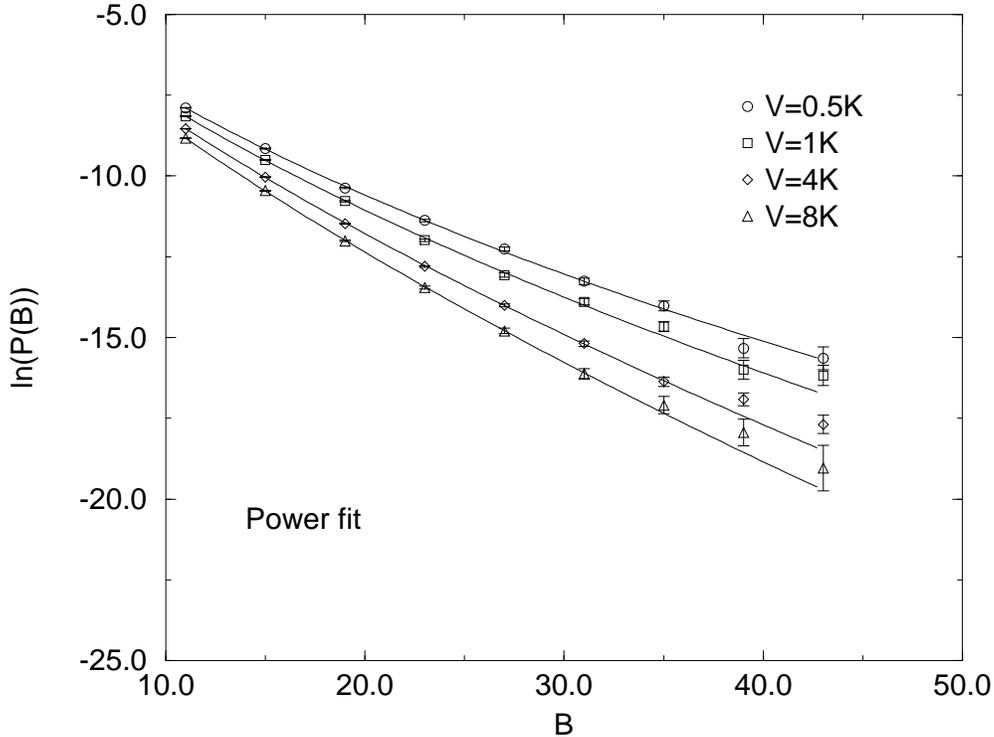}
\caption{$\log{P\left(B\right)}$ vs $B$ with a power fit
 assuming $\gamma=0.25$.}
\end{center}
\end{figure}

Figure 3 shows the same data now fitted according to the power
scenario Eq.\ (7). The best fit in this case yields
$a^\prime=-0.2(15)$, $\beta^\prime=-1.38(5)$ and $\delta=3(2)$ with
$\chi^2=6.3/6$ assuming $\gamma=0.25$ as before. 
At face value then it remains hard to differentiate between the
two situations. However, notice that the extracted value
of $\beta=0.056(1)$ from the log fit is more than twice its 
estimated value from the fits for the effective critical
coupling $\beta=0.025(1)$ (Table 1). 
In contrast the estimate for $\beta^\prime=-1.38(5)$ from
the power fit is quite close to its value estimated earlier
$\beta^\prime=-1.23(4)$. The relative proximity of the
two estimates is particularly impressive considering that
one is derived from the behaviour of baby universes with
size less than $8000$ simplices while the other is extracted
from the critical coupling at volumes much greater than $8000$.
Furthermore, it is clear that the power fit would still hold good
if we set $\delta=a^{\prime} =0$ so that such a fit (with a truly
minimal number of parameters) would do much better than the logarithm.

Additional information can be obtained by looking at the mean number of
nodes per unit volume. It is easy to see that this quantity
is related to the critical coupling through (see eg \cite{us3d})
\begin{equation}
\left<{N_0\over V}\right>={\partial\kappa_4^c\left(V,\kappa_0\right)\over
\partial\kappa_0}.
\end{equation}
Thus finite volume corrections to the effective critical coupling
result in similar finite volume corrections to $\left<N_0/V\right>$.
In Fig.\ 4 we show this quantity on a log-log scale together with
a least-square power fit. 
While the $\chi^2$ of such a fit is terrible, showing
that such a simple parameterization is insufficient to describe the
data in detail, the fit shows that a small power-law correction is
again rather well able to account for the overall structure
of the finite volume corrections\footnote{The fit yields
$\gamma=0.26$.}. Notice that any coefficient of
a logarithmic piece in $\kappa_4^c\left(V,\kappa_0\right)$ will
not contribute since it cannot depend on $\kappa_0$.

\begin{figure}
\begin{center}
\leavevmode
\epsfxsize=400pt
\epsfbox{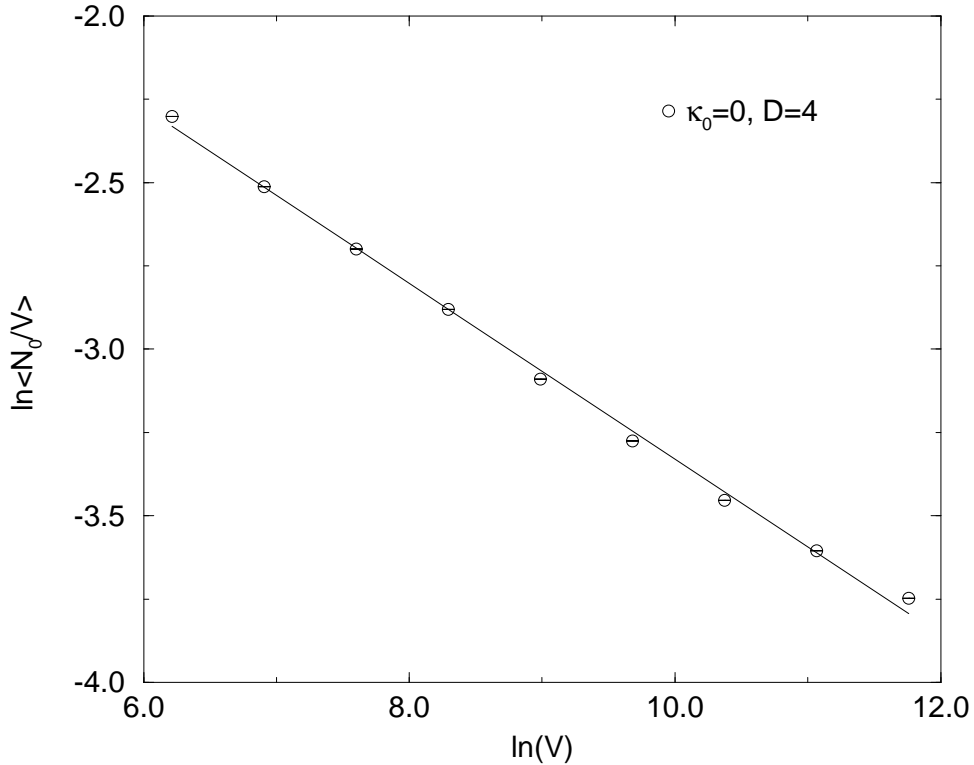}
\caption{Mean node number per unit volume with fit assuming
 a weak power convergence.}
\end{center}
\end{figure}
 
In conclusion, we have presented numerical results which, although not
definitive, are very consistent with the
existence of an exponential bound in the dynamical 
triangulation model of $4d$ quantum gravity. The evidence for this
comes both from fits to the volume dependence of the critical coupling, 
an analysis
of the baby universe distribution in the crumpled phase and the
scaling of the mean node number. 
Although individually these quantities are not very
conclusive, it is remarkable how consistent results are
obtained if we assume a weak power convergence.
Clearly, it is important
to strengthen these conclusions both by simulating intermediate
lattice volumes and perhaps via a high statistics simulation at
say volume $V=16000$ directed at probing further into the tail of
the baby universe distribution.

The calculations reported here were supported, in part, by grants
NSF PHY-9503371, PHY-9200148 and from research funds
provided by Syracuse University.

\vfill
\end{document}